\documentclass[twocolumn,A4]{article}
\usepackage[dvips]{graphics}

\begin{document}

\onecolumn

\begin{center}
{\bf{\Large Electron transport through molecular bridge systems}}\\
~\\
Santanu K. Maiti$^{1,2,*}$ \\
~\\
~\\
{\em $^1$Theoretical Condensed Matter Physics Division,
Saha Institute of Nuclear Physics, \\
1/AF, Bidhannagar, Kolkata-700 064, India \\
$^2$Department of Physics, Narasinha Dutt College,
129, Belilious Road, Howrah-711 101, India} \\
~\\
{\bf Abstract}
\end{center}
Electron transport characteristics are investigated through some molecular 
chains attached to two non-superconducting electrodes by the use of Green's 
function method. Here we do parametric calculations based on the tight-binding 
formulation to characterize the electron transport through such bridge 
systems. The transport properties are significantly influenced by 
(a) the length of the molecular chain and (b) the molecule-to-electrodes
coupling strength and here we focus are results in these aspects.
In this context we also discuss the steady state current fluctuations, 
the so-called shot noise, which is a consequence of the quantization 
of charge and is not directly available through conductance measurements.

\vskip 1cm
\begin{flushleft}
\end{flushleft}
\vskip 4.5in
\noindent
{\bf ~$^*$Corresponding Author}: Santanu K. Maiti

Electronic mail: santanu.maiti@saha.ac.in
\newpage
\twocolumn

\section{Introduction}

Recent progress in nanoscience and technology has allowed one to study 
the electron transport through molecules in a very controllable way. The 
prospect of individual devices based on single molecules is approaching
realization since electronic components are getting smaller and smaller.
Following experimental developments, theory can play a major role in
understanding the new mechanisms of conductance. The single-molecule 
electronics plays a key role in the design of future nanoelectronic 
circuits, but, the goal of developing a reliable molecular-electronics 
technology is still over the horizon and many key problems, such as 
device stability, reproducibility and the control of single-molecule 
transport need to be solved. Electronic transport through molecules 
was first studied theoretically in $1974$~\cite{aviram}. Since then 
numerous experiments~\cite{metz,fish,reed1,reed2,tali} have carried 
out on the electron transport through molecules placed between two 
metallic electrodes with few nanometer separation. It is very essential 
to control electron conduction through such quantum devices and the 
present understanding about it is quite limited. For example, it is 
not very clear how the molecular transport is affected by the structure 
of the molecule itself or by the nature of its coupling with the 
electrodes. The electron conduction through such two-terminal devices 
can be controlled by some bias voltage. The current passing across 
the junction then becomes a strongly non-linear function of the applied 
voltage, and its understanding is a highly challenging problem. The 
knowledge of current fluctuations (quantum origin) provides a key idea 
for fabrication of efficient molecular devices. Blanter {\em et 
al.}~\cite{butt} have described elaborately how the lowest possible 
noise power of the current fluctuations can be determined in a 
two-terminal conductor. The steady state current fluctuations, the 
so-called shot noise, is a consequence of the quantization of charge 
and it can be used to obtain information on a system which is not 
available through conductance measurements. The noise power of the 
current fluctuations provides an additional important information 
about the electron correlation by calculating the Fano factor ($F$) 
which directly informs us whether the magnitude of the shot noise 
achieves the Poisson limit ($F=1$) or the sub-Poisson ($F<1$) limit.

In the present article we provide a simple analytical formulation of 
the transport problem through some molecular chains of different lengths 
by using the tight-binding formulation. There exist several {\em ab initio} 
methods for the calculation of the conductance~\cite{yal,ven,xue,tay,der,
dam} as well as model calculations~\cite{muj1,muj2,sam,hjo,walc1,walc2,
baer3,baer1,baer2,tagami}. 
The model calculations are motivated by the fact that the {\em ab initio} 
methods are computationally too expensive, while, the model calculations 
by using the tight-binding formulation are computationally very cheap and 
also provide a worth insight to the problem. In our present study attention 
is drawn on the qualitative behavior of the physical quantities rather than 
the quantitative ones. Not only that, the {\em ab initio} theories do not 
give any new qualitative behavior for this particular study in which we 
concentrate ourselves.  

The article is specifically organized as follow. Following the introduction
(Section $1$), in Section $2$ we describe very briefly the technique for 
the calculation of the transmission probability ($T$), current ($I$) and 
the noise power its current fluctuations ($S$) through a molecular chain 
sandwiched between two metallic electrodes by means of the Green's function 
technique. In Section $3$, we study the behavior of the conductance as a 
function of the injecting electron energy, the current and the noise power 
of its fluctuations as a function of the applied bias voltage for some 
molecular chains in the aspects of (a) the length of the chain and (b) the 
molecule-to electrodes coupling strength. Finally, we draw our conclusions 
in Section $4$.

\section{The model and a brief description of the theoretical formulation}

\begin{figure}[ht]
{\centering \resizebox*{7cm}{1.15cm}{\includegraphics{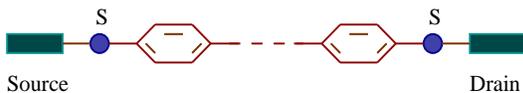}}\par}
\caption{(Color online). Schematic view of an array of benzene molecules 
attached to two metallic electrodes (made of gold in the actual 
experiment), the source and the drain, by thiol (sulfur-hydrogen i.e., 
S-H bond) groups (using the chemisorption technique).}
\label{benzene}
\end{figure}

In this section we describe very briefly about the methodology for the 
calculation of the transmission probability ($T$), conductance ($g$), 
current ($I$) and the noise power of current fluctuations ($S$) for 
a molecular chain sandwiched between two metallic electrodes 
(schematically represented as in Fig.~\ref{benzene}) by the use of 
Green's function technique.

At low temperature and bias voltage the conductance $g$ of the molecular 
chain is given by the Landauer conductance formula~\cite{datta},
\begin{equation}
g=\frac{2e^2}{h} T
\label{equ1}
\end{equation}
where the transmission probability $T$ is expressed as~\cite{datta},
\begin{equation}
T=Trace\left[\Gamma_S G_M^r \Gamma_D G_M^a\right]
\label{equ2}
\end{equation}
In this expression $G_M^r$ and $G_M^a$ are the retarded and advanced Green's 
functions of the molecular chain and $\Gamma_S$ and $\Gamma_D$ describe its 
coupling with the source and the drain, respectively. The Green's function 
of the molecular chain is written in this form,
\begin{equation}
G_M=\left(E-H_M-\Sigma_S-\Sigma_D\right)^{-1}
\label{equ3}
\end{equation} 
where $E$ is the energy of the injecting electron and $H_M$ is the 
Hamiltonian of the chain. This Hamiltonian can be expressed in the 
tight-binding form within the non-interacting picture like,
\begin{equation}
H_M=\sum_i \epsilon_i c_i^{\dagger} c_i + \sum_{<ij>} t \left(c_i^{\dagger} 
c_j + c_j^{\dagger} c_i\right)
\label{equ4}
\end{equation}
where $\epsilon_i$'s are the site energies and $t$ is the nearest-neighbor 
hopping strength. In Eq.(\ref{equ3}), $\Sigma_S$ and $\Sigma_D$ correspond to 
the self-energies due to the coupling of the molecular chain with the two 
electrodes. Now all the information about the electrode-to-molecule 
coupling are included into these two self-energies through the use of 
Newns-Anderson chemisorption theory and the detailed descriptions of this
theory can be found from the references~\cite{muj1,muj2}.  

The current passing across the chain can be assumed as a single electron
scattering process between the two reservoirs of charge carriers. The 
current-voltage relationship can be obtained from the 
expression~\cite{datta},
\begin{equation}
I(V)=\frac{e}{\pi \hbar}\int_{-\infty}^{\infty} \left(f_S-f_D\right) T(E) dE
\label{equ5}
\end{equation}
where the Fermi distribution function $f_{S(D)}=f\left(E-\mu_{S(D)}\right)$ 
with the electrochemical potentials $\mu_{S(D)}=E_F\pm eV/2$. For the sake 
of simplicity, here we assume that the entire voltage is dropped across the 
molecule-electrode interfaces and this assumption doesn't significantly 
affect the qualitative aspects of the current-voltage characteristics. This 
assumption is based on the fact that the electric field inside the molecular
chain especially for short chains, seems to have a minimal effect on the 
conductance-voltage characteristics. On the other hand for quite longer 
chains and higher bias voltage, the electric field inside the chain may 
play a more significant role depending on the internal structure of the 
chain~\cite{tian}, yet the effect is too small.

The noise power of the current fluctuations is calculated from the following
expression~\cite{butt},
\begin{eqnarray}
S & = & \frac{2e^2}{\pi \hbar}\int_{-\infty}^{\infty}\left[T(E)\left\{f_S
\left(1-f_S\right) + f_D\left(1-f_D\right) \right\} \right. \nonumber \\ 
 & & + T(E) \left. \left\{1-T(E)\right\}\left(f_S-f_D\right)^2 \right] dE
\label{equ6}
\end{eqnarray}
where the first two terms of this equation correspond to the equilibrium 
noise contribution and the last term gives the non-equilibrium or shot noise
contribution to the power spectrum. By calculating the noise power we can 
determine the Fano factor $F$, which is essential to predict whether the shot 
noise lies in the Poisson or the sub-Poisson limit, through the 
relation~\cite{butt},
\begin{equation}
F=\frac{S}{2 e I}
\label{equ7}
\end{equation}
The shot noise achieves the Poisson limit for $F=1$ and in this case no 
electron correlation exists between the charge carriers. On the other hand 
for $F<1$, the shot noise reaches the sub-Poisson limit and it provides the 
information about the electron correlation among the charge carriers.

Throughout this article we study our results only at absolute zero 
temperature, but the qualitative features of all the results are also
invariant up to some finite temperature ($\sim 300$ K). For simplicity, 
here we take the unit $c=e=h=1$ in our all calculations.

\section{Results and their interpretation}

Here we will study all the essential features of the electron transport 
characteristics for some molecular chains in the two distinct regimes. 
One is the so-called weak-coupling regime, which is defined as $\tau_{S(D)} 
<< t$ and the other one is the so-called strong-coupling regime, denoted as 
$\tau_{S(D)} \sim t$. The symbols $\tau_S$ and $\tau_D$ correspond to the 
hopping strengths of the molecular chain to the source and drain, 
respectively. In our present calculations, the parameters for these two 
regimes are chosen as $\tau_S=\tau_D=0.5$, $t=3$ (weak-coupling) and 
$\tau_S=\tau_D=2.5$, $t=3$ (strong-coupling).

\begin{figure}[ht]
{\centering \resizebox*{8.25cm}{9cm}{\includegraphics{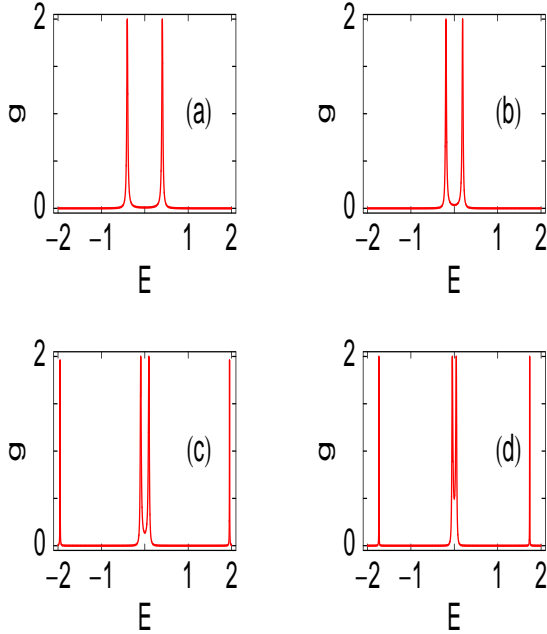}}\par}
\caption{(Color online). Conductance $g$ as a function of the injecting 
electron energy $E$ for the molecular bridges in the limit of weak 
molecular coupling. (a), (b), (c) and (d) correspond to the bridges 
with two, three, four and five benzene rings, respectively.}
\label{condlow}
\end{figure}

In Fig.~\ref{condlow} we plot the variation of the conductance $g$ as a 
function of the injecting electron energy $E$ for the molecular chains in 
the weak-coupling limit. Figures~\ref{condlow}(a), (b), (c) and (d) 
correspond to the results for the molecular chains with two, three, 
four and five benzene rings, respectively. The sharp resonant peaks 
are observed in the 
conductance spectra for some particular energy values, while for all other 
energies the peaks almost vanish. At these resonances the conductance $g$ 
approaches to $2$ indicating that the transmission probability $T$ goes to 
unity (since we get the relation $g=2T$ from the Landauer conductance 
formula, see Eq.(\ref{equ1}) with $e=h=1$ in our present formulation). 
These resonant peaks are associated with the molecular energy levels 
and their positions also change with the length of the molecular chain. 
Accordingly, we get more and more resonant peaks as we increase the 
length of the molecular chain. Thus it can be predicted that the 
conductance spectrum manifests itself the electronic structure of 
the molecular chain.

\begin{figure}[ht]
{\centering \resizebox*{8.25cm}{9cm}{\includegraphics{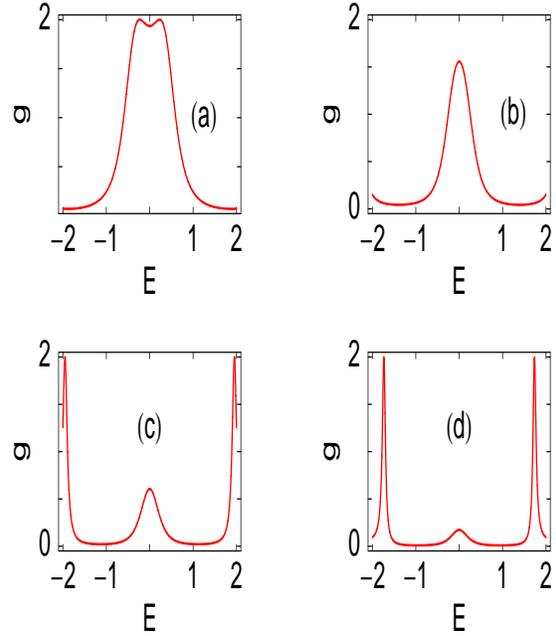}}\par}
\caption{(Color online). Conductance $g$ as a function of the injecting 
electron energy $E$ for the molecular bridges in the limit of strong 
molecular coupling. (a), (b), (c) and (d) correspond to the bridges with 
two, three, four and five benzene rings, respectively.}
\label{condhigh}
\end{figure}

In the limit of strong molecular coupling, the resonant peaks in the 
conductance spectra get substantial widths as shown in Fig.~\ref{condhigh}, 
where Figs.~\ref{condhigh}(a), (b), (c) and (d) correspond to the same 
molecular chains as in Fig.~\ref{condlow}. Such enhancement of the resonant 
widths is due to the broadening of the molecular energy levels caused by the 
coupling of the molecules to the side attached electrodes in this strong 
coupling limit, where the contribution comes from the imaginary parts of 
the two self-energies $\Sigma_S$ and $\Sigma_D$. 
From this figure (Fig.~\ref{condhigh}) two important features are observed
those are: (I) the width and (II) the height of the resonant peak across 
the energy $E=0$ gradually decrease with the increase of the length of the
molecular chain. Such kind of features are not clearly visible in the limit 
of weak molecular coupling (see Fig.~\ref{condlow}) since the widths of the
resonant peaks are much narrow for this coupling case. Now try to explain
these results. With the increase of the length of the molecular chain,
the two energy levels across the zero energy value (associated with the
two resonant peaks around the energy $E=0$, those are observed from
Fig.~\ref{condlow}) become closer and closer. This is due to the existence
of the molecular energy levels of the respective molecules at these energies
and they become broadened with the increase of the molecular coupling.
Since these two energy levels are very closely spaced they overlap with 
each other in some extent and produce a wide resonant peak across the
energy $E=0$, as shown in Fig.~\ref{condhigh}. The width of this
resonant peak decreases gradually with the length of the chain which 
is due to the closeness of the two energy levels around the energy $E=0$. 
Now the decrease of the height of the resonant peak across the energy 
$E=0$ is due to the interference effect of the waves traversing through 
the different arms of the molecular rings. For the chains with smaller 
number of rings, the number of interfering paths are small, while for 
the longer chains we get several paths. As a result of this interference 
effect, the probability amplitude of getting the electron across the 
chain becomes strengthened or weakened (from the standard interpretation 
of the quantum mechanical theory). The anti-resonances in the transmission 
(conductance) spectra are due to the cancellation of the transmittances 
along the two paths. Therefore, the probability amplitude $T$ becomes 
less than one and it gradually decreases with the increase of the length 
of the chain. Such anti-resonant states are specific to the interferometric 
nature of the scattering states and do not occur in the usual 
one-dimensional scattering problems involving potential barriers. From 
these results it can be emphasized that the electron transport 
characteristics significantly depend on the length of the molecular 
chain which provides a key idea for fabrication of molecular devices. 

\begin{figure}[ht]
{\centering \resizebox*{8.25cm}{9cm}{\includegraphics{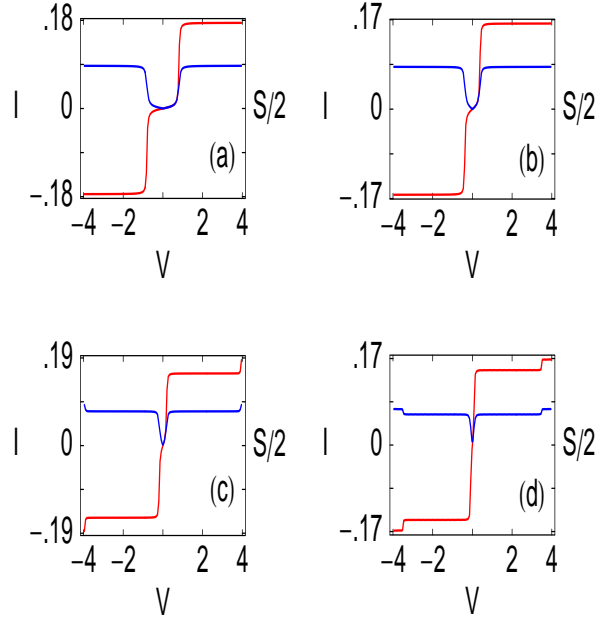}}\par}
\caption{(Color online). Current $I$ (red curve) and the noise power of 
its fluctuations $S$ (blue curve) as a function of the applied bias 
voltage $V$ for the molecular bridges in the weak molecule-to-electrodes 
coupling limit. (a), (b), (c) and (d) correspond to the bridges with 
two, three, four and five benzene rings, respectively.}
\label{currlow}
\end{figure}

Now we describe the behavior of the current $I$ and the noise power of its
fluctuations $S$ as a function of the applied bias voltage $V$ for these 
molecular bridges, where both the current and the noise power are evaluated 
through the integration procedure of the transmission function $T$ (see 
Eq.(\ref{equ5}) and Eq.(\ref{equ6})). In Fig.~\ref{currlow}, we plot the 
current and the noise power of its fluctuation for the molecular chains 
in the limit of weak-coupling, where Figs.~\ref{currlow}(a), (b), (c) 
and (d) correspond to the results for the chains with two, three, four and 
five benzene rings, respectively. The red and blue curves correspond to 
the current and the noise power, respectively. It is observed that the 
current shows staircase-like structure with sharp steps as a function of the
applied bias voltage. This is due to the existence of the sharp resonant 
peaks in the conductance spectra in the weak-coupling limit 
(Fig.~\ref{condlow}). As the bias voltage increases, the electrochemical 
potentials 
on the electrodes are gradually shifted and eventually cross one of the 
molecular energy levels. Accordingly, a current channel is opened and a jump 
in the $I$-$V$ curve appears. Another key observation is that for the
bridge with two benzene rings we get the non-zero value of the current 
after some critical value of the applied bias voltage (see the red curve of
Fig.~\ref{currlow}) i.e., the conduction of electron through the bridge is
started (on-state condition of the molecular bridge) beyond some critical 
value of the bias voltage. This threshold voltage of the electron conduction 
gradually decreases with the increase of the length of the molecular chain. 
Thus in the limit of weak molecular coupling one can tune the threshold 
bias voltage for electron conduction quite significantly by changing the 
length of the molecular chain. It is also noted that the current amplitudes 
for all such bridges are almost invariant since the widths of the resonant
peaks are nearly same for all these bridges (see Fig.~\ref{condlow}) in 
this coupling limit.

In the study of the noise power of the current fluctuations for the 
weak-coupling limit (blue curves of Fig.~\ref{currlow}) we see that for 
all these molecular wires the shot noise goes from the Poisson limit 
($F=1$) to the sub-Poisson limit ($F<1$) as long as we cross the first 
step in the current-voltage ($I$-$V$) characteristics. Therefore, we can 
predict that the electrons are correlated after the tunneling process has
completed. Here the electrons are correlated only in the sense that one 
electron feels the existence of the other according to the Pauli exclusion 
principle (since we have neglected all other electron-electron interactions 
in our present formalism). Another significant observation is that the 
amplitudes of the noise power in the sub-Poisson regime for all such wires 
are almost invariant. These results emphasize that in the weak-coupling 
limit the noise power of the current fluctuations within the sub-Poisson 
limit remains in the same level independent of the length of the molecular 
chain.

The behavior of the current and the noise power is also very interesting 
in the strong-coupling regime and in the forthcoming part we will describe
\begin{figure}[ht]
{\centering \resizebox*{8.25cm}{9cm}{\includegraphics{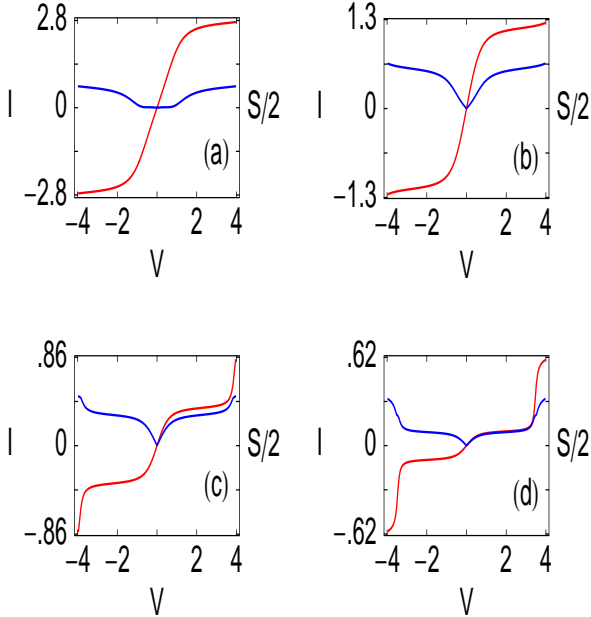}}\par}
\caption{(Color online). Current $I$ (red curve) and the noise power of 
its fluctuations $S$ (blue curve) as a function of the applied bias 
voltage $V$ for the molecular bridges in the strong molecule-to-electrodes 
coupling limit. (a), (b), (c) and (d) correspond to the bridges with two, 
three, four and five benzene rings, respectively.}
\label{currhigh}
\end{figure}
about that. In Fig.~\ref{currhigh} we display the results of the current 
and the noise power of its fluctuations in the strong-coupling limit for 
the molecular chains, where Figs.~\ref{currhigh}(a), (b), (c) and (d) 
correspond to the same systems as given in Fig.~\ref{currlow}. The red and 
blue curves represent the same meaning as in Fig.~\ref{currlow}. The shape 
and height of the current 
steps depend on the width of the resonances in the conductance spectra. With 
the increase of molecule-to-electrodes coupling strength, the step-like 
behavior disappears and the current becomes a continuous function of the 
applied bias voltage and its magnitude also increases (see the red curves of 
Fig.~\ref{currhigh}) compared to the weak-coupling case. Thus one can achieve
the larger current amplitude by tuning the molecular coupling strength, 
without changing any geometry of the molecular bridge. It is also noted that 
for the strong molecular coupling the current amplitude decreases with the 
increase of the number of benzene rings in the chain, which is clearly 
visible from the areas under the curves of the conductance spectra for 
the different bridges in this coupling limit (Fig.~\ref{condhigh}). 

Finally, in the study of the noise power of the current fluctuations we see 
that the noise power decreases (see the blue curves of Fig.~\ref{currhigh}) 
with the number of benzene rings in the chain for this strong molecular 
coupling. The significant observation is that for the chains with few number
of benzene rings (i.e., the chains with two, three and four benzene rings),
the current fluctuations (see Figs.~\ref{currhigh}(a), (b) and (c)) are in 
the sub-Poisson limit ($F<1$) and accordingly, there is no such possibility 
of transition from the Poisson ($F=1$) to the sub-Poisson limit, like as in 
the weak-coupling case. Therefore, the electron correlations are strongly
important for such chains (chains with few number of benzene rings) in this 
coupling limit. But for the longer chain i.e., the chain with five benzene 
rings the transition from the Poisson limit to the sub-Poisson limit takes
place beyond the first step in the current-voltage characteristic (see 
Fig.~\ref{currhigh}(d)). A similar kind of transition is also observed for 
the chains with more than five benzene rings in this strong-coupling limit 
which is not shown here in the figure. Thus we can emphasize that the noise 
power of the current fluctuations are less significant for the longer chains 
than that of the shorter ones in the limit of strong molecular coupling. 

\section{Concluding remarks}

In conclusion of this article, we have introduced a parametric approach 
based on the tight-binding model to investigate the electron transport 
characteristics through some molecular chains attached to two metallic 
electrodes. All the results studied here provide several key ideas for 
fabrication of efficient molecular devices.

For the weak-coupling limit, the conductance shows very sharp resonant peaks 
(Fig.~\ref{condlow}) associated with the molecular energy levels. The widths 
of these resonant peaks get broadened (Fig.~\ref{condhigh}) in the limit of 
strong molecular coupling, where the contribution comes from the imaginary 
parts of the two self-energies $\Sigma_S$ and $\Sigma_D$~\cite{datta}.

In the calculation of the current we have seen that the current shows 
staircase-like behavior (red curves in Fig.~\ref{currlow}) with sharp steps 
as a function of the applied bias voltage in the weak-coupling limit, while 
it gets continuous variation (red curves in Fig.~\ref{currhigh}) with the 
increase of the molecular coupling strength. For all such bridges the current 
amplitudes are comparable to each other in the weak molecular coupling, but
the current amplitude decays substantially with the increase of the length of 
the chain in the limit of strong-coupling.

Finally, in the determination of the noise power of the current fluctuations, 
we have noted that in the weak-coupling regime the current fluctuations 
within the sub-Poisson limit are in the same level (blue curves in
Fig.~\ref{currlow}) independent of the length of the molecular chain. On the 
other hand, for the strong-coupling regime the noise power of the current 
fluctuations are less important for the longer chains than that of the 
shorter ones (blue curves in Fig.~\ref{currhigh}).

Throughout our study, we have ignored the effect of electron-electron 
interaction and the influence of all the inelastic processes. More studies 
are expected to take into account the Schottky effect which comes from 
the charge transfer across the metal-molecule interfaces, the static 
Stark effect resulting from the modification of the electronic structure 
of the molecular bridge in presence of the applied bias voltage (essential 
especially for higher voltages).


\begin{thebibliography}{99}

\bibitem{aviram} A. Aviram and M. Ratner, Chem. Phys. Lett. \textbf{29},
277 (1974).
\bibitem{metz} R. M. Metzger, B. Chen, U. Hoepfner, M. V. Lakshmikantham,
D. Vuillaume, T. Kawai, X. Wu, H. Tachibana, T. V. Hughes, H. Sakurai,
J. W. Baldwin, C. Hosch, M. P. Cava, L. Brehmer, and G. J. Ashwell, J. Am. 
Chem. Soc. \textbf{119}, 10455 (1997).
\bibitem{fish} C. M. Fischer, M. Burghard, S. Roth, and K. V. Klitzing,
Appl. Phys. Lett. \textbf{66}, 3331 (1995).
\bibitem{reed1} J. Chen, M. A. Reed, A. M. Rawlett, and J. M. Tour, Science
\textbf{286}, 1550 (1999).
\bibitem{reed2} M. A. Reed, C. Zhou, C. J. Muller, T. P. Burgin, and J. M.
Tour, Science \textbf{278}, 252 (1997).
\bibitem{tali} T. Dadosh, Y. Gordin, R. Krahne, I. Khivrich, D. Mahalu, 
V. Frydman, J. Sperling, A. Yacoby, and I. Bar-Joseph, Nature \textbf{436},
677 (2005).
\bibitem{butt} Y. M. Blanter and M. B\"{u}ttiker, Phys. Rep. \textbf{336},
1 (2000).
\bibitem{yal} S. N. Yaliraki, A. E. Roitberg, C. Gonzalez, V. Mujica,
and M. A. Ratner, J. Chem. Phys. \textbf{111}, 6997 (1999).
\bibitem{ven} M. Di Ventra, S. T. Pantelides, and N. D. Lang, Phys. Rev.
Lett. \textbf{84}, 979 (2000).
\bibitem{xue} Y. Xue, S. Datta, and M. A. Ratner, J. Chem. Phys. 
\textbf{115}, 4292 (2001).
\bibitem{tay} J. Taylor, H. Gou, and J. Wang, Phys. Rev. B \textbf{63},
245407 (2001).
\bibitem{der} P. A. Derosa and J. M. Seminario, J. Phys. Chem. B 
\textbf{105}, 471 (2001).
\bibitem{dam} P. S. Damle, A. W. Ghosh, and S. Datta, Phys. Rev. B 
\textbf{64}, R201403 (2001).
\bibitem{muj1} V. Mujica, M. Kemp, and M. A. Ratner, J. Chem. Phys. 
\textbf{101}, 6849 (1994).
\bibitem{muj2} V. Mujica, M. Kemp, A. E. Roitberg, and M. A. Ratner, 
J. Chem. Phys. \textbf{104}, 7296 (1996).
\bibitem{sam} M. P. Samanta, W. Tian, S. Datta, J. I. Henderson, and
C. P. Kubiak, Phys. Rev. B \textbf{53}, R7626 (1996).
\bibitem{hjo} M. Hjort and S. Staftr\"{o}m, Phys. Rev. B \textbf{62}, 
5245 (2000).
\bibitem{walc1} K. Walczak, Phys. Stat. Sol. (b) \textbf{241}, 2555 (2004).
\bibitem{walc2} K. Walczak, Cent. Eur. J. Chem. \textbf{2}, 524 (2004).
\bibitem{baer3} D. Walter, D. Neuhauser, and R. Baer, Chem. Phys. 
\textbf{299}, 139 (2004).
\bibitem{baer1} R. Baer and D. Neuhauser, Chem. Phys. \textbf{281},
353 (2002).
\bibitem{baer2} R. Baer and D. Neuhauser, J. Am. Chem. Soc. \textbf{124},
4200 (2002).
\bibitem{tagami} K. Tagami, L. Wang, and M. Tsukada, Nano Lett. \textbf{4},
209 (2004).
\bibitem{datta} S. Datta, {\em Electronic transport in mesoscopic systems},
Cambridge University Press, Cambridge (1997).
\bibitem{tian} W. Tian, S. Datta, S. Hong, R. Reifenberger, J. I. Henderson,
and C. I. Kubiak, J. Chem. Phys. \textbf{109}, 2874 (1998).

\end{thebibliography}
\end{document}